\documentclass[11pt]{article}
\usepackage{hyperref}
\pdfoutput=1

\begin{document}

\title{Fluid Juggling}

\author{Enrique Soto$^1$ and Roberto Zenit$^2$ \\ $^1$Centro de Ciencias Aplicadas y Desarrollo Tecnol\'ogico\\ $^2$Instituto de Investigaciones en Materiales \\Universidad Nacional Aut\'onoma de M\'exico \\ Ciudad Universitaria, Coyoac\'an, 04510, M\'exico, D.F., M\'exico}
\maketitle

%% The abstract (in this file, and that submitted as text to arXiv) should include the exact phrase
%% "fluid dynamics video" or "fluid dynamics videos"

\begin{abstract}
This fluid dynamics video is an entry for the Gallery of Fluid Motion for the 66th Annual Meeting of the Fluid Dynamics Division of the American Physical Society. We show the curious behaviour of a light ball interacting with a liquid jet. For certain conditions, a ball can be suspended into a slightly inclined liquid jet. We studied this phenomenon using a high speed camera. The visualizations show that the object can be `juggled' for a variety of flow conditions. A simple calculation showed that the ball remains at a stable position due to a Bernoulli-like effect. The phenomenon is very stable and easy to reproduce.

\end{abstract}

% main text

\section{Introduction}
It is fascinating: when a liquid jet interacts with a ball, a kind of \emph{juggling} can be achieved. The ball can be trapped in air by its interaction with a fluid jet; the ball moves up and down around at a stable position. This experiment is very simple, everyone should try it. It is a simple demonstration of the beautiful flow behaviour of fluids.

The water jet impacts the bottom part of the ball; the jet turns into a flowing fluid film that surrounds the ball. Due to the change in thickness from the jet to the film, the mean film velocity is higher than that of the jet, resulting in a lower pressure region which in turn make the ball position stable. To successfully produce levitation, the liquid in the film needs to be evacuated (such that its weight does not break the force balance between the momentum flux of the jet and the weigh of the ball). This is the reason why the liquid jet has to be slightly inclined. The inclination and the flowing film induce rotation in the ball. The collision of the film with itself (flowing around the ball) produces an outgoing jet, which combined with the ball rotation and surface ripples, creates a beautiful series of films, threads and drops that are continuously ejected.

Some juggling tricks such as trap, catch, switch, rise up and swirl are shown in the video. We are eager to study this flow in more detail. To our knowledge, there has not been a methodical study of the flow shown here.

Two sample videos are
\href{http://ecommons.library.cornell.edu/bitstream/1813/8237/2/Juggling_hi_9oct.mp4}{Video
1, HR} and
\href{http://ecommons.library.cornell.edu/bitstream/1813/8237/4/Juggling_low_9oct.mpg}{Video
2, LR}.

\end{document}